\documentclass[aps,prd,a4paper,12pt,nofootinbib,superscriptaddress,oneside]{revtex4-1}
\usepackage[a4paper, margin=1in]{geometry}
\usepackage{epigraph}
\usepackage{amsmath,amssymb}
\usepackage{textcomp}
\usepackage{latexsym}
\usepackage{MnSymbol}
\usepackage{textcomp}
\usepackage{enumitem} 
\usepackage{bm}

\usepackage{graphicx}

\usepackage{color}
\definecolor{grey}{rgb}{0.6,0.6,0.6}
\definecolor{darkblue}{rgb}{0,0,.3}
\definecolor{darkgreen}{rgb}{0,.5,0}
\definecolor{deepred}{rgb}{0.4,0,0}
\usepackage[colorlinks,linkcolor=darkblue,urlcolor=deepred,citecolor=darkgreen]{hyperref}
\usepackage{bookmark}

\begin{document}
\title{A Thousand Problems in Cosmology\\ Chapter 4: Black holes}
\author{Yu.L.~Bolotin}
\email{ybolotin@gmail.com}
\affiliation{Kharkov Institute of Physics and Technology,\\
1 Akademicheskaya, Kharkov 61108, Ukraine}
\author{I. V. Tanatarov}
\email{igor.tanatarov@gmail.com}
\affiliation{Kharkov Institute of Physics and Technology,\\
1 Akademicheskaya, Kharkov 61108, Ukraine}
\affiliation{Department of Physics and Technology, Kharkov V.N. Karazin National
University, 4 Svoboda Square, Kharkov 61077, Ukraine}
\author{O. B. Zaslavskii}
\email{zaslav@ukr.net}
\affiliation{Department of Physics and Technology, Kharkov V.N. Karazin National
University, 4 Svoboda Square, Kharkov 61077, Ukraine}

\begin{abstract}
The fourth chapter of the collection of problems in cosmology, devoted to black holes. Consists of basic introduction, sections on Schwarzschild and Kerr black holes, a section on particles' motion and collisions in general black hole space-times, and the astrophysical part. This version contains only formulations of 137 problems. The full collection, with solutions included, is available in the form of a wiki-based resource at \href{www.universeinproblems.com}{www.universeinproblems.com}. The cosmological community is welcome to contribute to its development.
\end{abstract}

\maketitle

\tableofcontents
\newpage

\section{Technical warm-up}
The problems collected here introduce the tools needed for solving problems in the following sections. They can be solved in succession, or one can return to them if and when questions arise. For more background see the section ``Equations of General Relativity'' of chapter 2.

\subsection{Uniformly accelerated observer, Rindler metric}	
\emph{Einstein's equivalence principle states that locally a gravitational field cannot be dis\-tin\-guished from a non-inertial frame of reference. Therefore a number of effects of General Relativity, such as time dilation in a gravitational field and formation of horizons, can be studied in the frame of Special Theory of Relativity when considering uniformly accelerated observers.}
\begin{enumerate}
\item\label{BlackHole01}
Derive the equation of motion $x(t)$ of a charged particle in Minkowski space in a uniform electric field without initial velocity. Show that its acceleration is constant.
\item\label{BlackHole02}
What region of spacetime is unobservable for such an accelerated observer? In what region is this observer  unobservable?
\item\label{BlackHole03}
Consider the set of particles, which move with constant accelerations $a=const>0$ in Minkowski space, with initial conditions at time $t=0$ set as $x=\rho=c^{2}/a$. Let $\tau$ be the proper time of these particles in the units of $\rho/c$. What region of spacetime is parametrized by the pair of positive numbers $(\tau,\rho)$? Express the metric in this region in the coordinates $(\rho,\varphi)$, where $\varphi=c\tau/\rho$. This is the Rindler metric.
\end{enumerate}

\subsection{Metric in curved spacetime}
\emph{We see here, how, given an arbitrary metric tensor, to determine physical distance between points, local time and physical velocity of a particle in an arbitrary frame of reference.}

\emph{This problem, though fundamentally important,  is necessary in full form only for consideration of particle dynamics in the Kerr metric. In order to analyze the dynamics in the Schwarzschild metric, it suffices to answer all the questions with a substantially simplifying condition $g_{0i}=0$, where $i=1,2,3$ (see the last of the problems).}

Let the spacetime metric have the general form
\[ds^2=g_{\mu\nu}dx^{\mu}dx^{\nu}.\]
Coordinates are arbitrary and do not carry direct metrical meaning. An observer, stationary in a given coordinate frame, has 4-velocity $u^{\mu}=(u^{0},0,0,0)$, and the interval determines his proper ``local'' time
\[c^2 d\tau^{2}=ds^{2}=g_{00}(dx^{0})^2.\]
An observer in point A, with coordinates $x^\mu$, determines the physical ``radar'' distance to an infinitely close point $B$, with coordinates $x^{\mu}+dx^{\mu}$, in the following way. She sends a light beam to $B$ and measures the time it takes for the reflected beam to come back. Then distance to $B$ is half the proper time she waited from emission to detection times $c$. It is also natural for her to consider the event of the beam reflection in $B$ to be simultaneous with the middle of the infinitely small 4-distance between the events of emission and detection of light beam in $A$.

\begin{enumerate}[resume]
\item\label{BlackHole09}
Find the physical distance $dl$ between two events with coordinates $x^\mu$ and $x^\mu+dx^{\mu}$.
\item\label{BlackHole10}
Find the difference between coordinate times of two infinitely close simultaneous events.
\item\label{BlackHole11}
Let a particle's world line be $x^{\mu}(\lambda)$. What is the proper time interval $\delta\tau$ of a stationary observer, in which this particle covers distance from $x^{\mu}$ to $x^\mu+dx^\mu$?
\item\label{BlackHole12}
Physical velocity $v$ of the particle is defined as  $dl/\delta\tau$. Express it through the $4$-velocity of the particle and through its coordinate velocity $dx^{i}/dx^{0}$; find the interval along the world line $ds$ in terms of $v$ and local time $d\tau$.
\item\label{BlackHole13}
How are all the previous answers simplified if $g_{0i}=0$?
\end{enumerate}

\section{Schwarzschild black hole}
The spherically symmetric solution of Einstein's equations in vacuum for the spacetime metric has the form \cite{Schw}
\begin{align}\label{Schw}
ds^{2}=h(r)\,dt^2-h^{-1}(r)\,dr^2-r^2 d\Omega^{2},
	&\qquad\mbox{where}\quad
	h(r)=1-\frac{r_g}{r};\quad r_{g}=\frac{2GM}{c^{2}};\\
d\Omega^{2}=d\theta^{2}+\sin^{2}\theta\, d\varphi^{2}&\;\mbox{-- metric of unit sphere.}\nonumber
\end{align}
The Birkhoff's theorem (1923) \cite{Birkhoff,Jebsen} states, that this solution is unique up to coordinate transformations. The quantity $r_g$ is called the Schwarzschild radius, or gravitational radius, $M$ is the mass of the central body or black hole.

\subsection{Simple problems}
\begin{enumerate}[resume]
\item\label{BlackHole15}
Find the interval of local time (proper time of stationary observer) at a point $(r,\theta,\varphi)$ in terms of coordinate time $t$, and show that $t$ is the proper time of an observer at infinity. What happens when $r\to r_{g}$?

\item\label{BlackHole16}
What is the physical distance between two points with coordinates $(r_{1},\theta,\varphi)$ and $(r_{2},\theta,\varphi)$? Between $(r,\theta,\varphi_{1})$ and $(r,\theta,\varphi_{2})$? How do these distances behave in the limit $r_{1},r\to r_{g}$?

\item\label{BlackHole17}
What would be the answers to the previous two questions for $r<r_g$ and why\footnote{It is actually not a very simple problem}? Why the Schwarzschild metric cannot be imagined as a system of ``welded'' rigid rods in $r<r_g$, as it can be in the external region?

\item\label{BlackHole18}
Calculate the acceleration of a test particle with zero velocity.

\item\label{BlackHoleExtra1} 
Show that Schwarzschild metric is a solution to Einstein's equations.
\end{enumerate}

\subsection{Symmetries and integrals of motion of Schwarzschild metric}
\begin{enumerate}[resume]
\item\label{BlackHole19}
What integral of motion arises due to existance of a timelike Killing vector? Express it through the physical velocity of the particle.

\item\label{BlackHole20}
Derive the Killing vectors for a sphere in Cartesian coordinate system; in spherical coordinates.
\item\label{BlackHole21}
Verify that in coordinates $(t,r,\theta,\varphi)$ vectors 
\[ \begin{array}{l}
	\Omega^{\mu}=(1,0,0,0),\\
	R^{\mu}=(0,0,0,1),\\
	S^{\mu}=(0,0,\cos\varphi,-\cot\theta\sin\varphi),\\
	T^{\mu}=(0,0,-\sin\varphi,-\cot\theta\cos\varphi)
\end{array}\]
are the Killing vectors of the Schwarzschild metric.

\item\label{BlackHole22}
Show that existence of Killing vectors $S^\mu$ and $T^\mu$ leads to motion of particles in a plane.

\item\label{BlackHole23}
Show that the particles' motion in the plane is stable.

\item\label{BlackHole24}
Write down explicitly the conserved quantities  $p_{\mu}\Omega^{\mu}$ and $p_{\mu}R^{\mu}$ for movement in the plane $\theta=\pi/2$.

\item\label{BlackHole25}
What is the work needed to pull a particle from the horizon to infinity? Will a black hole's mass change if we drop a particle with zero initial velocity from immediate proximity of the horizon?

\end{enumerate}
\subsection{Radial motion in Schwarzchild metric}
Consider a particle's radial motion: $\dot{\varphi}=\dot{\theta}=0$. In this problem one is especially interested in asymptotes of all functions as $r\to r_{g}$.

\begin{enumerate}[resume]
\item\label{BlackHole26}
Derive the equation for null geodesics (worldlines of massless particles).

\item\label{BlackHole27}
Use energy conservation to derive $v(r)$, $\dot{r}(r)=dr/dt$, $r(t)$ for a massive particle. Initial conditions: $g_{00}|_{\dot{r}=0}=h_{0}$ (the limiting case $h_{0}\to 1$ is especially interesting and simple).

\item\label{BlackHole28}
Show that the equation of radial motion in terms of proper time of the particle is the same as in the non-relativistic Newtonian theory. Calculate the proper time of the fall from $r=r_0$ to the center. Derive the first correction in $r_{g}/r$ to the Newtonian result. Initial velocity is zero.

\item\label{BlackHole29}
Derive the equations of radial motion in the ultra-relativistic limit.

\item\label{BlackHole30}
A particle (observer) falling into a black hole is emitting photons, which are detected on the same radial line far away from the horizon (i.e. the photons travel from emitter to detector radially). Find  $r$, $v$ and $\dot{r}$ as functions of the signal's detection time in the limit  $r\to r_g$.
\end{enumerate}

\subsection{Blackness of black holes}
A source radiates photons of frequency $\omega_i$, its radial coordinate at the time of emission is $r=r_{em}$. Find the frequency of photons registered by a detector situated at $r=r_{det}$ on the same radial line in different situations described below. By stationary observers here, we mean stationary in the static Schwarzschild metric; ``radius'' is the radial coordinate $r$.
\begin{enumerate}[resume]
\item\label{BlackHole31}
The source and detector are stationary.

\item\label{BlackHole32}
The source is falling freely without initial velocity from radius $r_0$; it flies by the stationary detector at the moment of emission.

\item\label{BlackHole33}
The source is freely falling the same way, while the detector is stationary at $r_{det}>r_{em}$.

\item\label{BlackHole34}
The source is falling freely and emitting continuously photons with constant frequency, the detector is stationary far away from the horizon $r_{det}\gg r_{g}$. How does the detected light's intensity depend on $r_{em}$ at the moment of emission? On the proper time of detector?
\end{enumerate}

\subsection{Orbital motion, effective potential}
Due to high symmetry of the Schwarzschild metric, a particle's worldline is completely determined by the normalizing condition $u^{\mu}u_{\mu}=\epsilon$, where $\epsilon=1$ for a massive particle and $\epsilon=0$ for a massless one, plus two conservation laws---of energy and angular momentum.

\begin{enumerate}[resume]
\item\label{BlackHole35}
Show that the ratio of specific energy to specific angular momentum of a particle equals to $r_{g}/b$, where $b$ is the impact parameter at infinity (for unbounded motion).

\item\label{BlackHole36}
Derive the geodesics' equations; bring the equation for $r(\lambda)$ to the form
\[\frac{1}{2}\Big(\frac{dr}{d\lambda}\Big)^{2}
	+V_{\epsilon}(r)=\varepsilon,\]
where $V_{\epsilon}(r)$ is a function conventionally termed as effective potential.

\item\label{BlackHole37}
Plot and investigate the function $V(r)$. Find the radii of circular orbits and analyze their stability; find the regions of parameters $(E,L)$ corresponding to bound and unbound motion, fall into the black hole. Consider the cases of a) massless, b) massive particles.

\item\label{BlackHole38}
Derive the gravitational capture cross-section for a massless particle; the first correction to it for a massive particle ultra-relativistic at infinity. Find the cross-section for a non-relativistic particle to the first order in $v^2/c^2$.

\item\label{BlackHole39}
Find the minimal radius of stable circular orbit and its parameters. What is the maximum gravitational binding energy of a particle in the Schwarzschild spacetime?
\end{enumerate}

\subsection{Miscellaneous problems}
\begin{enumerate}[resume]
\item\label{BlackHole40}
Gravitational lensing is the effect of deflection of a light beam's (photon's) trajectory in the gravitational field. Derive the deflection of a photon's trajectory in Schwarzschild metric in the limit $L/r_{g}\gg 1$. Show that it is twice the value for a massive particle with velocity close to $c$ in the Newtonian theory.

\item\label{BlackHole41}
Show that the $4$-acceleration of a stationary particle in the Schwarzschild metric can be presented in the form
\[a_{\mu}=-\partial_{\mu}\Phi,\quad
	\text{where}\quad \Phi=\ln \sqrt{g_{00}}
		=\tfrac{1}{2}\ln g_{00}\]
is some generalization of the Newtonian gravitational potential.

\item\label{BlackHole42}
Let us reformulate the problem in a coordinate-independent manner. Suppose we have an arbitrary stationary metric with timelike Killing vector $\xi^\mu$, and we denote the $4$-velocity of a stationary observer by $u^{\mu}=\xi^{\mu}/V$. What is the $4$-force per unit mass that we need to apply to a test particle in order to make it stay stationary? Show in coordinate-independent way that the answer coincides with $\partial_{\mu}\Phi$ (up to the sign), and rewrite $\Phi$ in coordinate-independent form.

\item\label{BlackHole43}
Surface gravity $\kappa$ of the Schwarzschild horizon can be defined as acceleration of a stationary particle at the horizon, measured in the proper time of a stationary observer at infinity. Find $\kappa$.
\end{enumerate}

Solving Einstein's equations for a spherically symmetric metric of general form in vacuum (energy-momentum tensor equals to zero), one can reduce the metric to
\[ds^2=f(t)\Big(1-\frac{C}{r}\Big)dt^2
	-\Big(1-\frac{C}{r}\Big)^{-1}dr^2-r^2 d\Omega^2,\]
where $C$ is some integration constant, and $f(t)$ an arbitrary function of time $t$.

\begin{enumerate}[resume]
\item\label{BlackHole44}
Suppose all the matter is distributed around the center of symmetry, and its energy-momentum tensor is spherically symmetric, so that the form of $g_{\mu\nu}$ written above is correct. Show that the solution in the exterior region is reduced to the Schwarzschild metric and find the relation between $C$ and the system's mass $M$.

\item\label{BlackHole45}
Let there be a spherically symmetric void $r<r_{0}$ in the spherically symmetric matter distribution. Show that spacetime in the void is flat.

\item\label{BlackHole46}
Let the matter distribution be spherically symmetric and filling regions  $r<r_{0}$ and $r_{1}<r<r_{2}$ ($r_{0}<r_{1}$). Can one affirm, that the solution in the layer of empty space $r_{0}<r<r_{1}$ is also the Schwarzschild metric?
\end{enumerate}

\subsection{Different coordinates, maximal extension}
We saw that a particle's proper time of reaching the singularity is finite. However, the Schwarzschild metric has a (removable) coordinate singularity at $r=r_{g}$. In order to eliminate it and analyze the casual structure of the full solution, it is convenient to use other coordinate frames. Everywhere below we transform the coordinates $r$ and $t$, while leaving the angular part unchanged.

\begin{enumerate}[resume]
\item\label{BlackHole47}
Make coordinate transformation in the Schwarzschild metric near the horizon $(r-r_{g})\ll r_{g}$ by using physical distance to the horizon as a new radial coordinate instead of $r$, and show that in the new coordinates it reduces near the horizon to the Rindler metric.

\item\label{BlackHole48}
Derive the Schwarzschild metric in coordinates $t$ and  $r^\star=r+r_{g}\ln|r-r_g|$. How do the null geodesics falling to the center look like in $(t,r^\star)$? What range of values of $r^\star$ corresponds to the region $r>r_g$?

\item\label{BlackHole49}
Rewrite the metric in coordinates $r$ and $u=t-r^\star$, find the equations of null geodesics and the value of $g=det(g_{\mu\nu})$ at $r=r_{g}$. Likewise in coordinates $r$ and $v=t+r^\star$; in coordinates $(u,v)$. The coordinate frames $(v,r)$ and $(u,r)$ are called the ingoing and outgoing Eddington-Finkelstein coordinates.

\item\label{BlackHole50}
Rewrite the Schwarzschild metric in coordinates $(u',v')$ and in the Kruskal coordinates $(T,R)$ (Kruskal solution), defined as follows:
\[v'=e^{v/2r_g},\quad u'=-e^{-u/2r_g};\qquad
	T=\frac{u'+v'}{2},\quad R=\frac{v'-u'}{2}.\]
What are the equations of null geodesics, surfaces $r=const$ and $t=const$, of the horizon $r=r_{g}$, singularity $r=0$, in the coordinates $(T,R)$? What is the range space of $(T,R)$? Which regions in the Schwarzschild coordinates do the regions $\{\text{I}:\;R>|T|\}$, $\{\text{II}:\;T>|R|\}$, $\{\text{III}:\;R<-|T|\}$ and $\{\text{IV}:\;T<-|R|\}$ correspond to? Which of them are casually connected and which are not? What is the geometry of the spacelike slice $T=const$ and how does it evolve with time $T$?

\item\label{BlackHole51}
Pass to coordinates
\[v''=\arctan\frac{v'}{\sqrt{r_g}},\quad
	u''=\arctan\frac{u'}{\sqrt{r_g}}\]
and draw the spacetime diagram of the Kruskal solution in them.

\item\label{BlackHole52}
The Kruskal solution describes an eternal black hole. Suppose, for simplicity,  that some black hole is formed as a result of radial collapse of a spherically symmetric shell of massless particles. What part of the Kruskal solution will be realized, and what will not be? What is the casual structure of the resulting spacetime?
\end{enumerate}

\section{Kerr black hole}
Kerr solution is the solution of Einstein's equations in vacuum that describes a rotating black hole (or the metric outside of a rotating axially symmetric body) \cite{Kerr63}. In the Boyer-Lindquist coordinates \cite{BoyerLindquist67} it takes the form
\begin{align}\label{Kerr}
	&&ds^2=\bigg(1-\frac{2\mu r}{\rho^2}\bigg)dt^2
		+\frac{4\mu a \,r\sin^{2}\theta}{\rho^2}
				\;dt\,d\varphi
		-\frac{\rho^2}{\Delta}\;dr^2-\rho^2\, d\theta^2
	+\qquad\nonumber\\
	&&-\bigg(
		r^2+a^2+\frac{2\mu r\,a^2 \,\sin^{2}\theta}{\rho^2}
	\bigg) \sin^2 \theta\;d\varphi^2;\\
	\label{Kerr-RhoDelta}
&&\text{where}\quad
	\rho^2=r^2+a^2 \cos^2 \theta,\qquad
	\Delta=r^2-2\mu r+a^2.
\end{align}
Here $\mu$ is the black hole's mass, $J$ its angular momentum, $a=J/\mu$; $t$ and $\varphi$ are time and usual azimuth angle, while $r$ and $\theta$ are some coordinates that become the other two coordinates of the spherical coordinate system at $r\to\infty$.

\subsection{General axially symmetric metric}
\emph{A number of properties of the Kerr solution can be understood qualitatively without use of its specific form. In this problem we consider the axially symmetric metric of quite general kind}
\begin{equation}\label{AxiSimmMetric}
		ds^2=A dt^2-B(d\varphi-\omega dt)^{2}-
	C\,dr^2-D\,d\theta^{2},\end{equation}
\emph{where functions $A,B,C,D,\omega$ depend only on $r$ and $\theta$.}
\begin{enumerate}[resume]
\item\label{BlackHole53}
Find the components of metric tensor $g_{\mu\nu}$ and its inverse $g^{\mu\nu}$.
\item\label{BlackHole54}
Write down the integrals of motion corresponding to Killing vectors $\partial_t$ and $\partial_\varphi$.
\item\label{BlackHole55}
Find the coordinate angular velocity $\Omega=\tfrac{d\varphi}{dt}$ of a particle with zero angular momentum $u_{\mu}(\partial_{\varphi})^{\mu}=0$.
\item\label{BlackHole55plus}
Calculate $A,B,C,D,\omega$ for the Kerr metric.
\end{enumerate}

\subsection{Limiting cases}
\begin{enumerate}[resume]
\item\label{BlackHole56}
Show that in the limit $a\to 0$ the Kerr metric turns into Schwarzschild with $r_{g}=2\mu$.
\item\label{BlackHole57}
Show that in the limit $\mu\to 0$ the Kerr metric describes Minkowski space with the spatial part in coordinates that are related to Cartesian as
\begin{align*}
	&x=\sqrt{r^2+a^2}\;\sin\theta\cos\varphi,
		\nonumber\\
	&y=\sqrt{r^2+a^2}\;\sin\theta\sin\varphi,\\
	&z=r\;\cos\theta\nonumber,\\
	&\text{where}\quad
	r\in[0,\infty),\quad
	\theta\in[0,\pi],\quad \varphi\in[0,2\pi).\nonumber
\end{align*}
Find equations of surfaces $r=const$ and $\theta=const$ in coordinates $(x,y,z)$. What is the surface $r=0$?
\item\label{BlackHole58}
Write the Kerr metric in the limit $a/r \to 0$ up to linear terms.
\end{enumerate}

\subsection{Horizons and singularity}
Event horizon is a closed null surface. A null surface is a surface with null normal vector $n^\mu$:
\[n^{\mu}n_{\mu}=0.\]
This same notation means that $n^\mu$ belongs to the considered surface (which is not to be wondered at, as a null vector is always orthogonal to self). It can be shown further, that a null surface can be divided into a set of null geodesics. Thus the light cone touches it in each point: the future light cone turns out to be on one side of the surface and the past cone on the other side. This means that world lines of particles, directed in the future, can only cross the null surface in one direction, and the latter works as a one-way valve, -- ``event horizon''.
\begin{enumerate}[resume]
\item\label{BlackHole59}
Show that if a surface is defined by equation $f(r)=0$, and on it $g^{rr}=0$, it is a null surface.
\item\label{BlackHole60}
Find the surfaces $g^{rr}=0$ for the Kerr metric. Are they spheres?
\item\label{BlackHole61}
Calculate surface areas of the outer and inner horizons.
\item\label{BlackHole62}
What values of $a$ lead to existence of horizons?
\end{enumerate}
On calculating curvature invariants, one can see they are regular on the horizons and diverge only at $\rho^2 \to 0$. Thus only the latter surface is a genuine singularity.
\begin{enumerate}[resume]
\item\label{BlackHole63} 
Derive the internal metric of the surface $r=0$ in Kerr solution.
\item\label{BlackHole64} 
Show that the set of points $\rho=0$ is a circle. How it it situated relative to the inner horizon?
\end{enumerate}

\subsection{Stationary limit}
Stationary limit is a surface that delimits areas in which particles can be stationary and those in which they cannot. An infinite redshift surface is a surface such that a phonon emitted on it escapes to infinity with frequency tending to zero (and thus its redshift tends to infinity). The event horizon of the Schwarzschild solution is both a stationary limit and an infinite redshift surface (see problems \ref{BlackHole31}-\ref{BlackHole34}). In the general case the two do not necessarily have to coincide.
\begin{enumerate}[resume]
\item\label{BlackHole65}
Find the equations of surfaces $g_{tt}=0$ for the Kerr metric. How are they situated relative to the horizons? Are they spheres?
\item\label{BlackHole66}
Calculate the coordinate angular velocity of a massless particle moving along $\varphi$ in the general axially symmetric metric  (\ref{AxiSimmMetric}). There should be two solutions, corresponding to light traveling in two opposite directions. Show that both solutions have the same sign on the surface $g_{tt}=0$. What does it mean? Show that on the horizon $g^{rr}=0$ the two solutions merge into one. Which one?
\item\label{BlackHole67}
What values of angular velocity can be realized for a massive particle? In what region angular velocity cannot be zero? What can it be equal to near the horizon?
\item\label{BlackHole68}
A stationary source radiates light of frequency $\omega_{em}$. What frequency will a stationary detector register? What happens if the source is close to the surface $g_{tt}=0$? What happens if the detector is close to this surface?
\end{enumerate}

\subsection{Ergosphere and the Penrose process}
Ergosphere is the area between the outer stationary limit and the outer horizon. As it lies before the horizon, a particle can enter it and escape back to infinity, but $g_{tt}<0$ there. This leads to the possibility of a particle's energy in ergosphere to be also negative, which leads in turn to interesting effects.

\textit{All we need to know of the Kerr solution in this problem is that it \emph{has an ergosphere}, i.e. the outer horizon lies beyond the outer static limit, and that on the external side of the horizon all the parameters $A,B,C,D,\omega$ are positive (you can check!). Otherwise, it is enough to consider the axially symmetric metric of general form.}
\begin{enumerate}[resume]
\item\label{BlackHole69}
Let a massive particle move along the azimuth angle $\varphi$, with fixed $r$ and $\theta$. Express the first integral of motion $u_t$ through the second one\footnote{Relations problem \ref{BlackHole12} and  \ref{BlackHole19}) do not hold, as they were derived in assumption that $g_{00}>0$.} $u_{\varphi}$ (tip: use the normalizing condition $u^\mu u_{\mu}=1$).
\item\label{BlackHole70}
Under what condition a particle can have $u_{t}<0$? In what area can it be fulfilled? Can such a particle escape to infinity?
\item\label{BlackHole71}
What is the meaning of negative energy? Why in this case (and in GR in general) energy is \emph{not} defined up to an additive constant?
\item\label{BlackHole72}
Let a particle $A$ fall into the ergosphere, decay into two particles $B$ and $C$ there, and particle $C$ escape to infinity. Suppose $C$'s energy turns out to be greater than $A$'s. Find the bounds on energy and angular momentum carried by the particle $B$.
\end{enumerate}

\subsection{Integrals of motion}
\begin{enumerate}[resume]
\item\label{BlackHole73}
Find the integrals of motion for a massless particle moving along the azimuth angle $\varphi$ (i.e. $dr=d\theta=0$). What signs of energy $E$ and angular momentum $L$ are possible for particles in the exterior region and in ergosphere?
\item\label{BlackHole74}
Calculate the same integrals for massive particles. Derive the condition for negativity of energy in terms of its angular velocity $d\varphi/dt$. In what region can it be fulfilled? Show that it is equivalent to the condition on angular momentum found in problem \ref{BlackHole70}.
\item\label{BlackHole75}
Derive the integrals of motion for particles with arbitrary $4$-velocity $u^{\mu}$. What is the allowed interval of angular velocities $\Omega=d\varphi/dt$? Show that for any particle $(E-\tilde{\Omega} L )>0$ for any $\tilde{\Omega}\in(\Omega_{-},\Omega_{+})$.
\end{enumerate}

\subsection{The laws of mechanics of black holes}
If a Killing vector is null on some null hypersurface $\Sigma$, $\Sigma$ is called a Killing horizon.
\begin{enumerate}[resume]
\item\label{BlackHole76} 
Show that vector  $K=\partial_{t}+\Omega_{H}\partial_{\varphi}$ is a Killing vector for the Kerr solution, and it is null on the outer horizon $r=r_{+}$. Here $\Omega_{H}=\omega\big|_{r=r_+}$ is the angular velocity of the horizon.
\item\label{BlackHole77}
Let us define the surface gravity for the Kerr black hole as the limit
\[\kappa=\lim\limits_{r\to r_{+}}
	\frac{\sqrt{a^{\mu}a_{\mu}}}{u^0}\]
for a particle near the horizon with $4$-velocity $\bm{u}=u^{t}(\partial_{t}+\omega\partial_{\varphi})$. In the particular case of Schwarzschild metric this definition reduces to the one given in problem \ref{BlackHole43}. Calculate $\kappa$ for particles with zero angular momentum in the Kerr metric. What is it for the critical black hole, with $a=\mu$?
\item\label{BlackHole78}
Find the change of (outer) horizon area of a black hole when a particle with energy $E$ and angular momentum $L$ falls into it. Show that it is always positive.
\item\label{BlackHole79}
Let us define the irreducible mass $M_{irr}$ of Kerr black hole as the mass of Schwarzschild black hole with the same horizon area. Find $M_{irr}(\mu,J)$ and $\mu(M_{irr},J)$. Which part of the total mass of a black hole can be extracted from it with the help of Penrose process?
\item\label{BlackHole80}
Show that an underextremal Kerr black hole (with $a<\mu$) cannot be turned into the extremal one in any continuous accretion process.
\end{enumerate}
This problem's results can be presented in the form that provides far-reaching analogy with the laws of thermodynamics.
\begin{itemize}
\item[0:] Surface gravity $\kappa$ is constant on the horizon of a stationary black hole. The zeroth law of thermodynamics: a system in thermodynamic equilibrium has constant temperature $T$.
\item[1:] The relation
\[\delta\mu=\frac{\kappa}{8\pi}\delta A_{+}
	+\Omega_{H}\delta J\]
gives an analogy of the first law of thermodynamics, energy conservation.
\item[2:] Horizon area $A_+$ is nondecreasing. This analogy with the second law of thermodynamics hints at a correspondence between the horizon area and entropy.
\item[3:] There exists no such continuous process, which can lead as a result to zero surface gravity. This is an analogy to the third law of thermodynamics: absolute zero is unreachable.
\end{itemize}

\subsection{Particles' motion in the equatorial plane}
The following questions refer to a particle's motion in the equatorial plane $\theta=\pi/2$ of the Kerr metric.
\begin{enumerate}[resume]
\item\label{BlackHole81}
Put down explicit expressions for the metric components and parameters $A,B,C,D,\omega$.
\item\label{BlackHole82}
What is the angular velocity of a particle with zero energy?
\item\label{BlackHole83}
Use the normalizing conditions for the $4$-velocity $u^{\mu}u_{\mu}=\epsilon^2$ and two conservation laws to derive geodesic equations for particles, determine the effective potential for radial motion.
\item\label{BlackHole84}
Integrate the equations of motion for null geodesics with $L=aE$, investigate the asymptotes close to the horizons, limits $a\to 0$ and $a\to \mu$.
\item\label{BlackHole85}
Find the minimal radii of circular geodesics for massless particles, the corresponding values of integrals of motion and angular velocities. Show that of the three solutions one lies beyond the horizon, one describes motion in positive direction and one in negative direction. Explore the limiting cases of Schwarzschild $a\to0$ and extreme Kerr $a\to\mu$.
\item\label{BlackHole86}
Find $L^2$ and $E^2$ as functions of radii for circular geodesics of the massive particles.
\item\label{BlackHole87}
Derive equation for the minimal radius of a stable circular orbit; find the energy and angular momentum of a particle on it, the minimal radius in the limiting cases $a/\mu\to 0,1$.
\end{enumerate}


\section{Physics in general black hole spacetimes}
In this section we use the $(-+++)$ signature, Greek letters for spacetime indices and Latin letters for spatial indices.

\subsection{Frames, time intervals and distances}
In the next several problems we again consider the procedure of measuring time and space intervals by different observers, but in a different, more formal and powerful approach.

\begin{enumerate}[resume]
\item\label{OZ01} Let a particle move with the four-velocity $U^{\mu }$. It can be viewed as
some observer carrying a frame attached to him. Locally, it defines the
hypersurface orthogonal to it. Show that 
\begin{equation}
h_{\mu \nu }=g_{\mu \nu }+U_{\mu }U_{\nu }  \label{h}
\end{equation}
is (i) the projection operator onto this hypersurface, and at the same time
(ii) the induced metric of the hypersurface. This means that  (i) for any
vector projected at this hypersurface by means of $h^\mu_\nu$, only the components orthogonal to $U^{\mu}$ survive, (ii) the repeated application of the projection operation leaves the vector within the hypersurface unchanged. In other words, $h_{\mu \nu}$ satisfies 
\begin{align}
&h^{\mu}_{\nu}U^{\nu }=0 ;  \label{1} \\
&h^{\mu}_{\nu }h^{\nu}_{\lambda}=h^{\mu }_{\lambda}.  \label{2}
\end{align}


\item\label{OZ02} 

Let us consider a particle moving with the four-velocity $U^{\mu }$. The
interval $ds^{2}$ between two close events is defined in terms of
differentials of coordinates,%
\begin{equation}
ds^{2}=g_{\mu \nu }dx^{\mu }dx^{\nu }.
\end{equation}

For given $dx^{\mu }$, what is the value of the proper time $d\tau _{obs}$ between the corresponding events measured by this observer? How can one define
locally the notions of simultaneity and proper distance $dl$ for the observer in terms of its four-velocity and the corresponding projection operator $h^\mu_\nu$ ? How is the interval $ds^{2}$ related to $d\tau _{obs}$ and $dl$?

\item\label{OZ03} 

Let our observer measure the velocity of some other particle passing in its
immediate vicinity. Relate the interval to $d\tau _{obs}$ and the particle's
velocity $w$.

\item\label{OZ04}

Analyze the formulas derived in the previous three problems applied to the case of flat spacetime (Minkovskii space) and compare them to
the known formulas of special relativity.

\item\label{OZ05}

Consider an observer being at rest with respect to a given coordinate frame: 
$x^{i}=const$ ($i=1,2,3$). Find $h_{\mu \nu }$, $d\tau _{obs}$, the
condition of simultaneity and $dl^{2}$ for this case. Show that the
corresponding formulas are equivalent to eqs. (84.6), (84.7) of \cite{LL},
where they are derived in a different way.

\item\label{OZ06}

Consider two events at the same point of space but at different values of
time. Find the relation between $dx^{\mu}$ and $d\tau _{obs}$ for such an
observer.
\end{enumerate}

\subsection{Fiducial observers}
\begin{enumerate}[resume]

\item\label{OZ07}

Consider an observer with 
\begin{equation}
U_{\mu }=-N\delta _{\mu }^{0}=-N(1,0,0,0)\text{.}  \label{uz}
\end{equation}%
We call it a fiducial observer (FidO) in accordance with \cite{mb}. This
notion is applied in \cite{mb} mainly to static or axially symmetric rotating
black holes. In the latter case it is usually called the ZAMO (zero angular
momentum observer). We will use FidO in a more general context. 

Show that a FidO's world-line is orthogonal to hypersurfaces of constant time $t=const$.

\item\label{OZ08}

Find the explicit form of the metric coefficients in terms of the components
of the FidO's four-velocity. Analyze the specific case of axially symmetric metric in
coordinates $(t,\phi ,r,\theta )$ with $g_{0i}=g_{t\phi }\delta _{i}^{\phi }$.

\item\label{OZ09}

Consider a stationary metric with the time-like Killing vector field $\xi^\mu =(1,0,0,0)$. Relate the energy $E$ of a particle with four-velocity $u^\mu$ as measured at
infinity by a stationary observer to that measured by a local observer with 4-velocity $U^\mu$.

\item\label{OZ10}

Express $E_{rel}$ and $E$ in terms of the relative velocity $w$ between a
particle and the observer (i.e. velocity of the particle in the frame of the observer and vice versa).

\item\label{OZ11}

Show that in the flat spacetime the formulas derived in the previous problem are reduced to the usual ones
of the Lorentz transformation.

\item\label{OZ12}

Find the expression for $E$ for the case of a static observer ($U^{i}=0$).

\item\label{OZ13}

Find the expression for $E$ for the case of the ZAMO observer and, in particular, in case of axially symmetric metric.

\end{enumerate}

\subsection{Collision of particles: general relationships}

\begin{enumerate}[resume]

\item\label{OZ14}

Let two particles collide. Define the energy in the center of mass (CM)
frame $E_{c.m.}$ at the point of collision and relate it to $E_{rel}$ and
the Lorentz factor of relative motion of the two particles.

\item\label{OZ15}

Let us consider a collision of particles 1 and 2 viewed from the frame
attached to some other particle 0. How are different Lorentz factors related
to each other? Analyze the case when the laboratory frame coincides with
that of particle 0.

\item\label{OZ16}

When can the relative Lorentz factor of two particles $\gamma $ as a function of their individual Lorentz factors in some frame $\gamma _{1}$ and $\gamma _{2}$ grow
unbounded? How can the answer be interpreted in terms of relative velocities?

\item\label{OZ17}

A tetrad basis, or the orthonormal tetrad, is the set of four unit vectors $h_{(a)}^\mu$ (subscripts in parenthesis $a=0,1,2,3$ enumerate these vectors), of which one, $h_{(0)}^\mu$, is timelike, and three vectors $h_{(i)}^\mu$ ($i=1,2,3$) are spacelike, so that
\begin{equation}
g_{\mu\nu}h_{(a)}^\mu h_{(b)}^\nu =\eta_{ab},\qquad a,b=0,1,2,3.
\end{equation}
A vector's tetrad components are
\begin{equation}
u_{(a)}=u_\mu h_{(a)}^\mu,\qquad u^{(b)}=\eta^{ab}u_{(b)}.
\end{equation}

Define the local three-velocities with the help of the tetrad basis attached to the observer, which
would generalize the corresponding formulas of special relativity.

\item\label{OZ18}

Derive the analogues of the results of problems \ref{OZ09} and \ref{OZ10} for massless particles (photons). Analyze the cases of static and ZAMO observers.

\item\label{OZ19}

The ergosphere is a surface defined by equation $g_{00}=0$. Show that it is
the surface of infinite redshift for an (almost) static observer.

\item\label{OZ20}

Consider an observer orbiting with a constant angular velocity $\Omega $ in
the equatorial plane of the axially symmetric back hole. Analyze what
happens to redshift when the angular velocity approaches the minimum or
maximum values $\Omega _{\pm }$.

\item\label{OZ21}

Let two massive particle 1 and 2 collide. Express the energy of each particle
in the centre of mass (CM) frame in terms of their relative Lorentz factor $\gamma (1,2)$. Analyze the limiting cases of ultra-relativistic $\gamma
(1,2)\rightarrow \infty $ and non-relativistic $\gamma (1,2)\approx 1$ collisions.

\item\label{OZ22}

For a stationary observer in a stationary space-time the quantity $\alpha =(U^0)^{-1}$ is the redshifting factor: if this observer emits a ptoton with frequency $\omega_{em}$, it is detected at infinity by another stationary observer with frequency $\omega_{det}=\alpha \omega_{em}$. For a generic observer this interpretation is invalid, however, $\alpha =ds/dt$ still determines the time dilation for this observer, and thus can still be called the same way. Express the redshifting factor of the center of mass frame $\alpha _{c.m.}$ through the redshifting factors of the colliding particles $\alpha_{1}$ and $\alpha_2$. 

\item\label{OZ23}

Relate the energy of a particle at infinity $E_{1}$, its energy at the point of collision in the C.M. frame $(E_{1})_{c.m.}$ and $\mu$.

\item\label{OZ24}

Solve the same problem when both particles are massless (photons). Write
down formulas for the ZAMO observer and for the C.M. frame.
\end{enumerate}

\section{Astrophysical black holes}

\subsection{Preliminary}
\begin{enumerate}[resume]

 \item\label{BlackHole88} Calculate in the frame of Newtonian mechanics the time of collapse of a uniformly distributed spherical mass with density $\rho_0$.

 \item\label{BlackHole89} Why are stars of a certain type called ``white dwarfs''?

  \item\label{BlackHole90} What is the physical reason for the stopping of thermonuclear reactions in the stars of white dwarf type?

  \item\label{BlackHole91} Estimate the radius and mass of a white dwarf.

  \item\label{BlackHole92} What is the average density of a white dwarf of one solar mass, luminosity one thousandth of solar luminosity and surface temperature twice that of the Sun?

  \item\label{BlackHole93} Explain the mechanism of explosion of massive enough white dwarfs, with masses close to the Chandrasekhar limit.

\item\label{BlackHole94} Thermonuclear explosions of white dwarfs with masses close to the Chandrasekhar limit lead to the phenomenon of supernova explosions of type I. Those have lines of helium and other relatively heavy elements in the spectrum, but no hydrogen lines. Why is that?

\item\label{BlackHole95} A supernova explosion of type II is related to the gravitational collapse of a neutron star. There are powerful hydrogen lines in their spectrum. Why?

\item\label{BlackHole96} Estimate the radius and mass of a neutron star.

\item\label{BlackHole97}	 Why do neutron stars have to possess strong magnetic fields?

\item\label{BlackHole98} Find the maximum redshift of a spectral line emitted from the surface of a neutron star.

 \item\label{BlackHole99} What is the gravitational radius of the Universe? Compare it with the size of the observable Universe.

 \item\label{BlackHole100} What is the time (the Salpeter time) needed for a black hole, radiating at its Eddington limit, to radiate away all of its mass?

\item\label{BlackHole101} The time scales of radiation variability of active galactic nuclei (AGNs) are from several days to several years. Estimate the linear sizes of AGNs.

\item\label{BlackHole102} What mechanisms can be responsible for the supermassive black hole (SMBH) in the center of a galaxy to acquire angular momentum?

\item\label{BlackHole103} The Galactic Center is so ``close'' to us, that one can discern individual stars there and examine  in detail their movement. Thus, observations carried out in 1992-2002 allowed one to reconstruct the orbit of motion of one of the stars (S2) around the hypothetical SMBH at the galactic center of the Milky Way. The parameters of the orbit are: period $15.2$ years, maximum distance from the black hole $120$ a.u., eccentricity $0.87$. Using this data, estimate the mass of the black hole.

\item\label{BlackHole104} Using the results of the previous problem, determine the density of the SMBH at the Galactic Center.

\item\label{BlackHole105} Show that for a black hole of mass $M$ the temperature of the surrounding hot gas in thermal equilibrium is proportional to
\[T\sim {{M}^{-1/4}}.\]

\item\label{BlackHole106} Show that luminosity of a compact object (neutron star or black hole) of several solar masses is mostly realized in the X-rays.

\item\label{BlackHole107} In order to remain bound while subject to the rebound from gigantic radiative power,
AGNs should have masses $M>{{10}^{6}}{{M}_{\odot }}$. Make estimates.

\item\label{BlackHole108} AGNs remain active for more than tens of millions of years. They must have tremendous masses to maintain the luminosity
\[L\sim {{10}^{47}}\text{erg/sec}\]
during such periods. Make estimates for the mass of an AGN.

\item\label{BlackHole109} What maximum energy can be released at the merger of two black holes with masses ${{M}_{1}}={{M}_{2}}=\frac{M}{2}$?

\item\label{BlackHole110} Show that it is impossible to divide a black hole into two black holes.

\item\label{BlackHole111} J.~Wheeler noticed that in the frame of classical theory of gravity the existence of black holes itself contradicts the law of entropy's increase. Why is that?

\item\label{BlackHole112} What is the reason we cannot attribute the observed entropy's decrease (see the previous problem) to the interior of the black hole?

\item\label{BlackHole113} Find the surface area of a stationary black hole as a function of its parameters: mass, angular momentum and charge.
\end{enumerate}

\subsection{Quantum effects}
\begin{enumerate}[resume]

\item\label{BlackHoleQ1} Estimate the maximum density of an astrophysical black hole, taking into account that black holes with masses $M<{{10}^{15}}g$ would not have lived to our time due to the quantum mechanism of evaporation.

\item\label{BlackHoleQ2} Determine the lifetime of a black hole with respect to thermal radiation.

\item\label{BlackHoleQ3} Determine the temperature of a black hole (Hawking temperature) of one solar mass, and the temperature of the supermassive black hole at the center of our Galaxy.

\item\label{BlackHoleQ4} Particles and antiparticles of given mass $m$ (neutrinos, electrons and so on) can be emitted only if the mass $M$ of the black hole is less than some critical mass ${M}_{cr}$. Estimate the critical mass of a black hole $M_{cr}(m)$.

\end{enumerate}


\begin{thebibliography}{99}
\bibitem{Schw} K. Schwarzschild, On the gravitational field of a mass point according to Einstein's theory, \textit{Sitzungsber. Preuss. Akad. Wiss. Phys. Math. Kl.},  p.189 (1916) [\href{http://arxiv.org/abs/physics/9905030v1}{arXiv:physics/9905030v1}].

\bibitem{Birkhoff} G.D. Birkhoff, Relativity and Modern Physics, p.253, Harvard University Press, Cambridge (1923).

\bibitem{Jebsen} J.T. Jebsen, \textit{Ark. Mat. Ast. Fys.} (Stockholm) \textbf{15}, nr.18 (1921), see also \href{http://arxiv.org/abs/physics/0508163}{arXiv:physics/0508163v2}.

\bibitem{Kerr63} R.P. Kerr, Gravitational field of a spinning mass as an example of algebraically special metrics. \textit{Phys. Rev. Lett.} \textbf{11} (5), 237 (1963).

\bibitem{BoyerLindquist67} R.H. Boyer, R.W. Lindquist. Maximal Analytic Extension of the Kerr Metric. \textit{J. Math. Phys} \textbf{8}, 265вЂ“281 (1967).

\bibitem{Wald} Wald R.M, General relativity. U. Chicago, 1984, 505p (ISBN 0226870332).

\bibitem{Carroll} Carroll S., Spacetime and geometry: an introduction to General Relativity. AW, 2003, 525p (ISBN 0805387323).


\bibitem{LL} Landau L.D., Lifshitz E.M. Vol. 2. The classical theory of fields [4ed., Butterworth-Heinemann, 1994].

\bibitem{t'Hooft} G. `t Hooft, Introduction to General Relativity, Caputcollege 1998.

\bibitem{MTW} 
Charles W. Misner, Kip S. Thorne. John Archibald Wheeler,  Gravitation. W.H. Freeman and Company, 1973.


\bibitem{Hobson2006} Hobson M., Efstathiou G., Lasenby A. General relativity: an introduction for physicists. CUP, 2006 (ISBN 0521536391).

\bibitem{Padmanabhan2010} Padmanabhan T. Gravitation: Foundations and Frontiers. CUP, 2010 (ISBN 9780521882231).

\bibitem{li} A. P. Lightman, W. H. Press, R. H. Price, and S. A. Teukolsky, Problem book in Relativity and Gravitation (Princeton University Press, Princeton, New Jersey, 1975).

\bibitem{mb} Black Holes: The Membrane Paradigm. Edited by Kip S. Thorne, Richard H. Price, Douglas A. Macdonald. Yale University Press New Haven and London, 1986.
\end{thebibliography}
\end{document}